\documentclass[12pt]{article}
\usepackage{graphicx}
\usepackage{amsmath}
\usepackage{amssymb}

\begin{document}

\bibliographystyle{hplain}

%
\title{Capivara: A decentralized package version control using Blockchain}

%
\author{Felipe Zimmerle da N. Costa and Ruy J. Guerra B. de Queiroz}

\begin{abstract}
Distributed consensus and Blockchains are popular among the cryptocurrencies where no one except the coins users, owns the data and transactions. No different to open source repositories, where the data belongs to the users. In this work it is presented a manner of having a repository for software packages in a Blockchain with distributed consensus, supported by the idea of the also demonstrated proof-of-download.
\end{abstract}

%
\maketitle

\emergencystretch 20em

\section{Introduction}
A repository for packages is part of every Linux distribution. Having a repository is a manner to ensure easy update and installation for the system and general end-user application packages. Software repositories are also utilized in Mobile Phone application stores. The same model is also adopted to search/download libraries for script languages or even browser extensions.

On retail stores the packages are only published given the discretion of the store. The store is in charge to ensure the authenticity of the package, sometimes the integrity or even the quality. Also, it guarantees the license fee. Moreover, it protects against unauthorized use of the software.

Although the objective of every store is pretty similar, a package store for an open source software platform has its nuances. Guaranteeing the quality, authenticity and integrity are also requirements for the majority of the distribution. However, transparency is a fundamental piece. The packages are validated using web-of-trust \cite{caronni2000walking}, allowing the packages to be distributed in different mirrors (trusted or not) as it is going to be also validated on the endpoint where the installation is conditioned to validation.

Another facet that is very common on open source software is the support for third-party repositories. Generally supplied from a small group of developers/users specifically to peers that demands, for whatever the reason, a more frequently updated or even customized package. Those packages may or may not be trusted by the distribution, but trusted by the end-user.

The central authority is presented in the software repository by the figure of the distribution, which is placed in that role not only due to its responsibility to vouch for the packages but also due to technical limitations on the way the packages are shared and distributed. These technical limitations could be circumvented with techniques already known and used in Blockchains where a {\it peer-to-peer} network with no central authority manages to have a consensus.

Here we wish to present a schema to implement a decentralized package repository using Blockchain. As for the experiments, the focus will be given on the open source repositories, given the easy access and accountability to the data of the package.

{\it This work is divided in the following sections}: A introduction on Blockchain (Section \ref{chp:blockchain}); Brief introduction on Distributed Consensus (Section \ref{chp:mining}); Software repositories (Section \ref{chp:repositories}); On Section \ref{chp:problem}
the problem is better illustrated, while Section \ref{chp:proposal} presents  a proposal followed by the Experiment (Section \ref{chp:experiment}) and finally the results (Section \ref{chp:results}) which is superseded by the conclusions (Section \ref{chp:conclusion}).


\section{Blockchain}
\label{chp:blockchain}

\begin{figure}[!ht]
  \centering
  \includegraphics[scale=0.30]{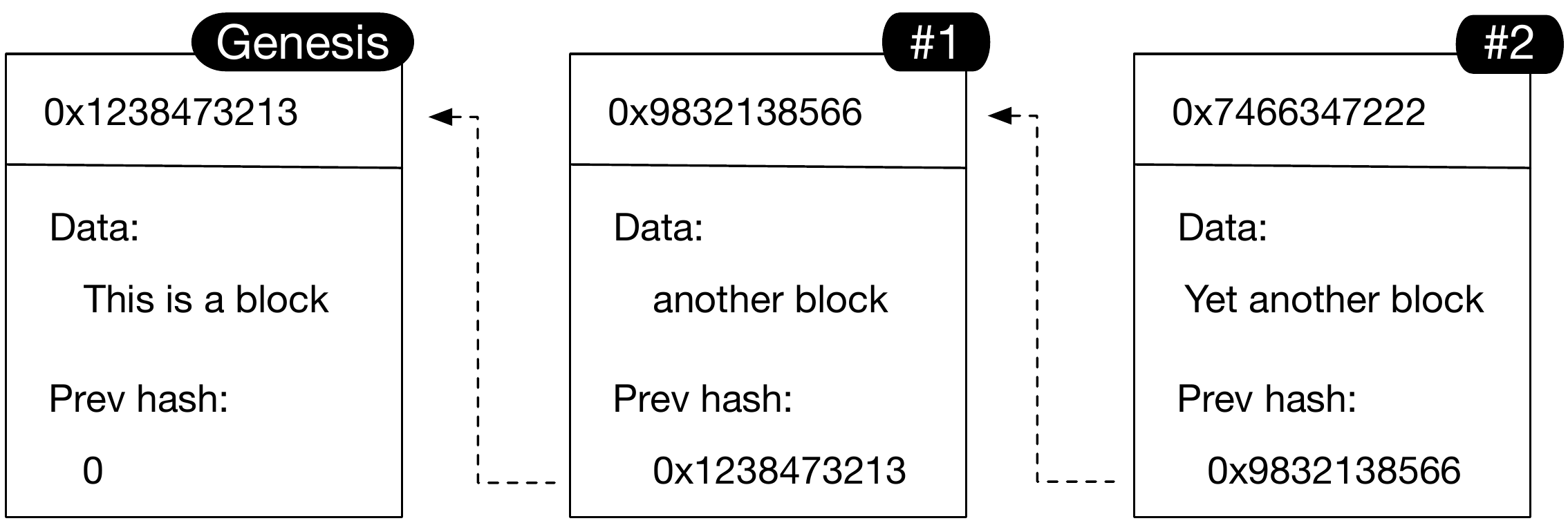}
  \caption{Blockchain blocks structure. Every block contains the hash of its predecessor.}
  \label{fig:blockchain-structure}
\end{figure}

Blockchain consists in a sequence of blocks that always contains the cryptographic hash of the previews block. The block hash took into consideration the previous block hash. Therefore, validating one is also to validate its predecessor, till the \textit{genesis} block. The \textit{genesis} Block is the first block of a Blockchain. The connections between the different blocks in a Blockchain are illustrated on \textbf{Figure \ref{fig:blockchain-structure}}.

The blocks are generated given a time beat, every N units of time a new block is likely to be generated. The blocks do not only contain their meta data but a payload. The payload is the data that is meaningful to the final application, regardless of the Blockchain structure. In the case of Bitcoin, the payload is a Merkle tree containing all the transactions held in the given block. A Merkle tree is a data structure that allows a rapid and inexpensive check of Bitcoin transactions \cite{okupski2014bitcoin}.

\begin{figure}[!ht]
  \centering
  \includegraphics[scale=0.45]{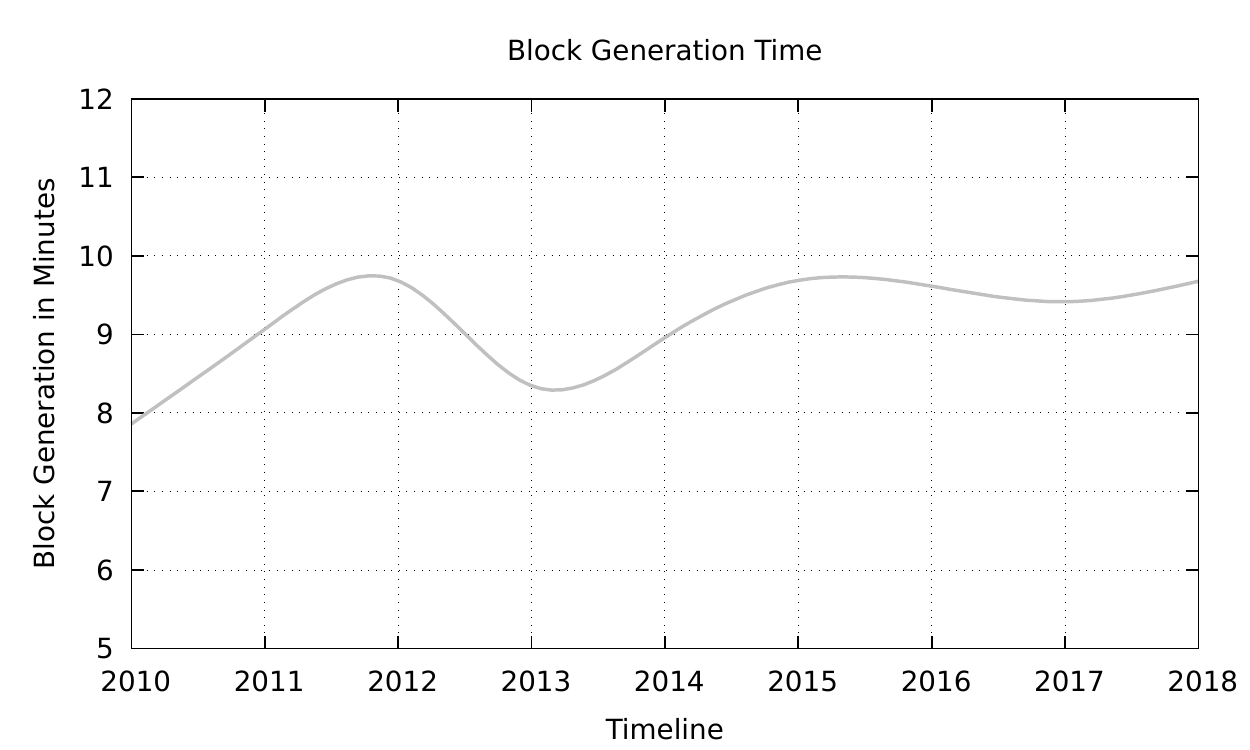}
  \caption{Amount of time in between Bitcoin blocks generation \cite{databct}.}
  \label{fig:block-generation-time}
\end{figure}

The time in between the blocks generation is not precise, as the generation of a new block relies on the network consensus, which may be a resolution of a cryptographic puzzle, explained on \textbf{Section \ref{chp:mining}}. On \textbf{Figure \ref{fig:block-generation-time}} it is possible to see the historical values of the amount of time that was necessary to generate Bitcoin blocks \cite{databct}. In Bitcoin, the target is to have each block generated in every 10 minutes \cite{nakamoto2008bitcoin}.

In Bitcoin the puzzle is adjusted according to the network power, the difficulty is adjusted on every 2016 blocks (about two weeks). The difficult is illustrated at \textbf{Equation  \ref{eq:block_difficult}}, where N is the new difficult, O is the old difficult and T is the time that took to compute the last 2016 blocks.

\begin{equation}\label{eq:block_difficult}
\centering
N=O*(2016*10)/T
\end{equation}

One of the biggest advantages of the Blockchain is its decentralized nature. It is possible to trust the Blockchain regardless of the channel of download and the source. As by design, it is resistant to modification of the data. It is "an open, distributed ledger that can record transactions between two parties efficiently and in a verifiable and permanent way". \cite{BlockChainTruth}

Once recorded, the data in any given block cannot be changed unless the subsequent blocks are also changed. 

\begin{figure}[!ht]
  \centering
  \includegraphics[scale=0.24]{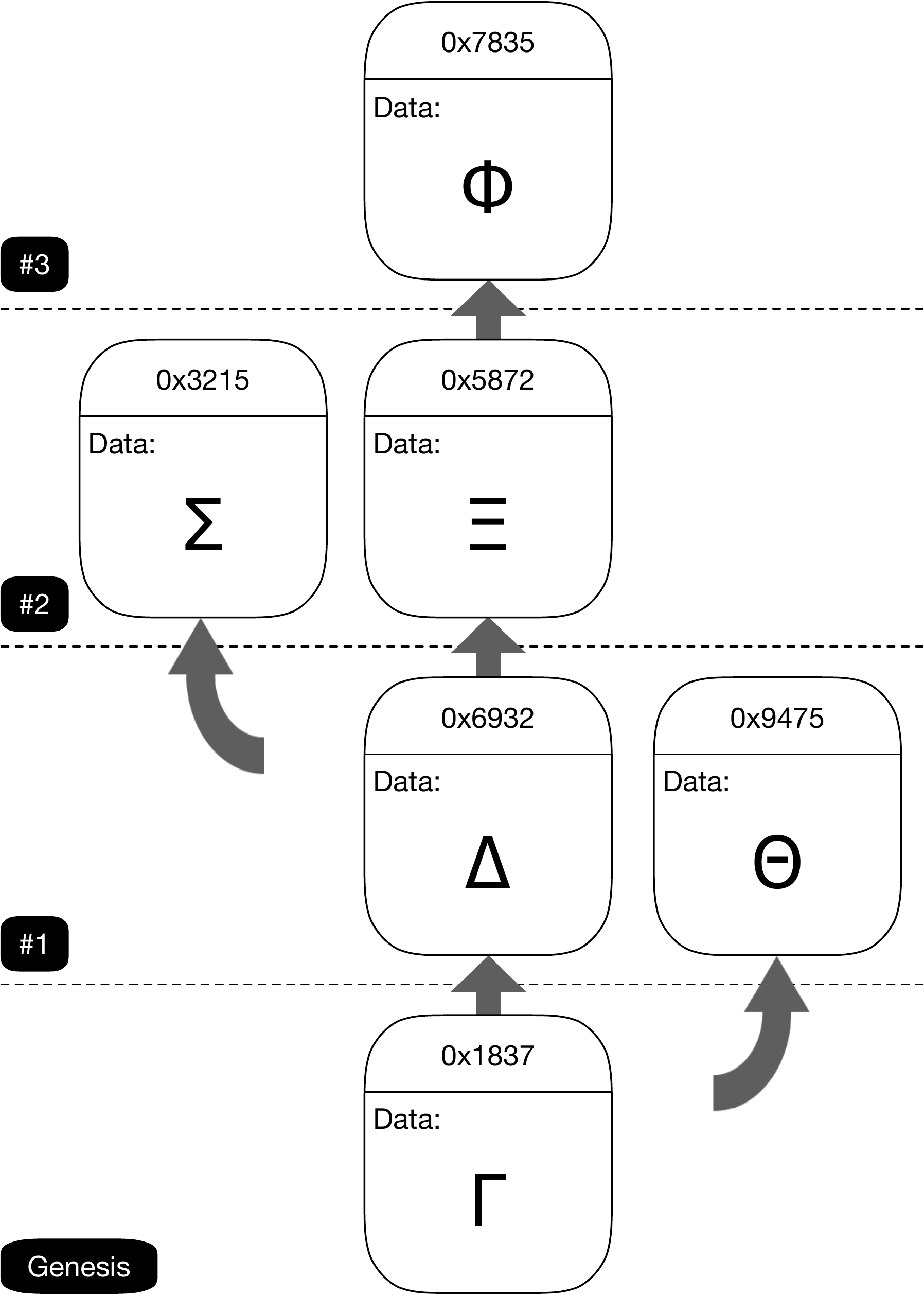}
  \caption{Hypothetical example of a Blockchain diverging at height 1 and 2; Consensus found on 3.}
  \label{fig:blockchain-diverged}
\end{figure}

Although the blocks have a single parent, it is possible that a block has multiple children. Each of those child blocks refers to and contains the hash of the same previews block. This scenario happens on a \textbf{Blockchain fork}, which is when blocks are generated almost concurrently from two (or more) different peers. In the latter case, it is a temporary situation. A diverging chain is illustrated on \textbf{Figure \ref{fig:blockchain-diverged}}.

The most popular use of the Blockchains are in cryptocurrencies \footnote{cryptocurrency: Digital asset used as a medium of exchange which secures the transactions using strong encryption as the example of the Bitcoin.}. However, it is feasible to be used beyond that, in applications that are not cryptocurrencies related.

\subsection{Structure}

The validation of the Blockchain is relatively inexpensive \cite{andreas2014mastering}. In order to be validated, the hash of the block in question needs to be computed. If it matches with a parent hash, every subsequent block is also validated. 

A Blockchain is decentralized using a \textit{peer-to-peer} network, where the participants are authenticated by their collaboration power, mitigating the possibility of infinite reproducibility, intrinsic to a digital asset \cite{andreas2014mastering}.

In a Blockchain, there are two different ways to refer to a block: \textbf{Block Hash} and \textbf{Block Height}. The block hash identifies the block uniquely and unambiguously. The hash is a digital fingerprint of the block while the height is a number that is given after the size of the Blockchain while the block was generated — being 0 the genesis block.

\subsubsection{Block}

In cryptocurrencies the block holds a Merkle tree \cite{nakamoto2008bitcoin} that contains the transactions. Among of the payload, the block also includes metadata: time stamping, the previous block hash, version of software/protocol, nonce and difficult target and the Merkle root. \textbf{Table \ref{tab:blockanatomy}} describes the fields utilized in a Bitcoin block.

\begin{table}
\centering
\begin{tabular}{ l|l }
 \hline
 Field & Description\\
 \hline \hline
 Version   & Version number of software/protocol.\\
 Previous Block Hash   & The hash of the previous block.\\
 Timestamp   & The time of the block creation.\\
 Difficult Target   & The proof-of-work difficult target.\\
 Nonce   & Number used on the block generation.\\
 Merkle root & The hash of the Merkle-Tree root.\\
 \hline
\end{tabular}
 \caption{\label{tab:blockanatomy}The anatomy of a block.}
\end{table}

In case blocks are produced concurrently by different nodes, there is an specific algorithm that will decide which chain is the valid one given the chain history.

There is no guarantee that any particular entry will remain in the best version of the history forever. As the incentives are made to extend new blocks, in opposite to overwrite, it is unlikely that old blocks became orphan. Orphan is the block that are not selected for inclusion in the main chain.

%
%

\subsubsection{Hard forks}

Whenever there is a change in the manner to generate new blocks, it is called \textbf{Hard Fork}. A hard fork happens when the old method of validation can no longer validate new blocks.

The \textbf{Hard Fork} is an essential mechanism to implement new features and fixes in a Blockchain. If the nodes are not in agreement with the hard fork, there will be a split where the Blockchain will assume two different consensuses one for each version. Therefore two different Blockchains.

\section{Distributed Consensus}
\label{chp:mining}

In a Blockchain there are different ways to generate a new block, depending on the consensus algorithm utilized by the Blockchain's \textit{peer-to-peer} network. The consensus mechanism needs to ensure that the decision upon the last block is acceptable to all legit nodes as it will define the truth of the Blockchain. This process of generating a new block is referred to as \textbf{mining}.

A valuable property that is in general likely to have in a Blockchain is the capability to not concentrate the power into a selected peer or peers, but make the network autonomous, without a central authority. Generally speaking, the consensus mechanism should be secure enough that it is more profitable to cooperate than it is to subvert.

As already demonstrated sometimes the consensus is not explicitly found in a given height, but later it is established because overall is less cost-effective to overwrite than to compute a new block.

The \textit{Proof-of-Work} is the most famous method of authentication/mining. It is the model adopted in the Bitcoin protocol \cite{nakamoto2008bitcoin}.

\subsubsection{Proof-of-Work (PoW)} In a Bitcoin network, a new block can be generated by any member of the network. However, in order to generate this block, the member has to solve a cryptographic puzzle.

The \textit{Proof-of-Work} puzzle relies on a computation of a hash with particular characteristics, which are adjusted considering the possibility to find the solution to the puzzle. The puzzle is expected to be solved giving a target: the amount of time that is expected in between the blocks generation.

In order to encourage peers to participate in the mining business, there is an incentive to the ones that generate the new block. The first transaction in the block is a payment to the user who has generated the block. This incentive is paid out in the format of tokens, which latter can be exchanged for money — the number of tokens halves on every 210,000 blocks. Eventually, the value will be zero, and this incentive will no longer exist. There is also an incentive to validate transactions.

In a Blockchain, there is no central authority. The consensus in the next block is giving by simple rules that every node should follow. A mined block that has not followed the rules is considered to be invalid, therefore likely to be an orphan block. In Blockchain, there is no election or fixed moment where the consensus occurs. Instead, the consensus is achieved of the asynchronous interaction of thousands of independent nodes, all following simple rules.

In order to better understand the \textit{Proof-of-work} it is first necessary to understand cryptographic hashes. Those hashes are the representation of input data of any size into a short string. There are, however, some characteristics that are important in a cryptographic hash, including: Unfeasible to manipulate a given content to generate a specific hash which will be considered valid by the verification mechanism.

\begin{figure}[!ht]
  \centering
  \includegraphics[scale=0.340]{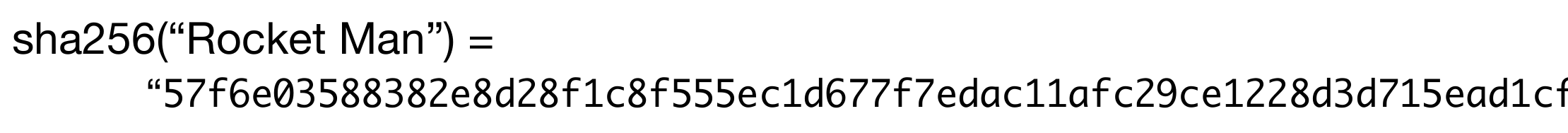}
  \caption{Example of the sha256 output for the input "Rocket Man"}
  \label{fig:hash}
\end{figure}

\begin{figure}[!ht]
  \centering
  \includegraphics[scale=0.340]{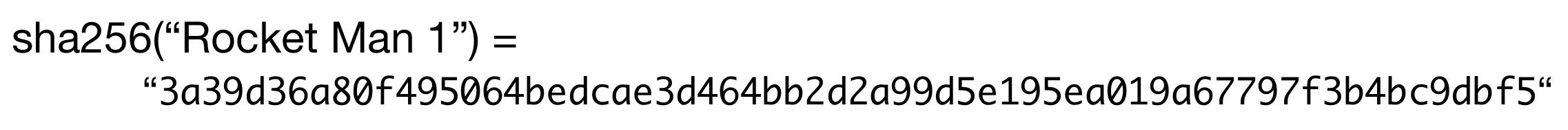}
  \caption{Example of the sha256 output for the input "Rocket Man" added the \textit{nonce} 1.}
  \label{fig:hash2}
\end{figure}

In \textit{sha256}, a popular cryptographic hash algorithm, for any input given, the output is a set of 256 bits. Virtually impossible to find two inputs that lead to the same hash output. The \textbf{Figure \ref{fig:hash}} exemplifies the output of a \textit{sha256} hash. By changing the phrase a little bit, we will have a completely different output. As demonstrated at \textbf{Figure \ref{fig:hash2}}.

In \textbf{Figure \ref{fig:hash2}} it was added a \textit{nonce} to the string illustrated at \textbf{Figure \ref{fig:hash}}. The \textit{nonce} is commonly used in cryptography to avoid common hash output due to very common input. With different  \textit{nonce}, the hash will be most likely different.

\begin{figure}[!ht]
  \centering
  \includegraphics[scale=0.340]{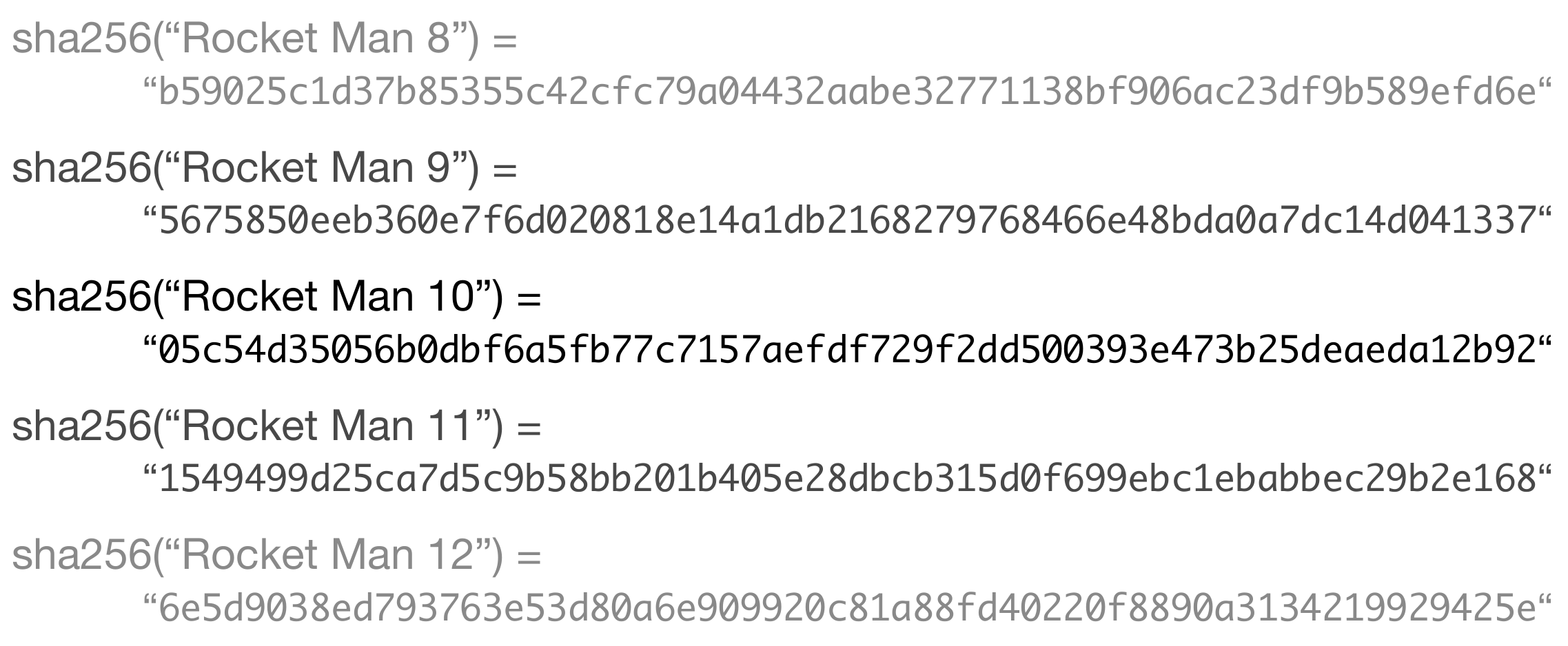}
  \caption{Example of the sha256 output for the input "Rocket Man" added the \textit{nonce} 10.}
  \label{fig:hash3}
\end{figure}

To illustrate a puzzle, imagine if there is a need to find a hash where the first character is zero. Given the data from the \textbf{Figure \ref{fig:hash}} and \textbf{Figure \ref{fig:hash2}}, it is possible to change the \textit{nonce}, until finding a hash that starts with zero. That is a relatively easy task. It is possible to notice that the \textit{nonce} 10 will generate the hash output of \textbf{Figure \ref{fig:hash3}}.

Considering the hash is evenly distributed, it is expected to have a hash starting with 0 once every 16 hashes. In numerical terms, that is the same to find a hash that has a value less than
{\small
0x1000000000000000000000000000000000000000000000000000000000000000}. That is an example of a Bitcoin challenge target.

Notice that the smaller the target is, the more difficult it is to find a hash that has a value smaller than it. Once the \textit{nonce} is found, anyone can rapidly and inexpensively validate that the \textit{nonce} meets the target.

\begin{figure}[!ht]
  \centering
  \includegraphics[scale=0.40]{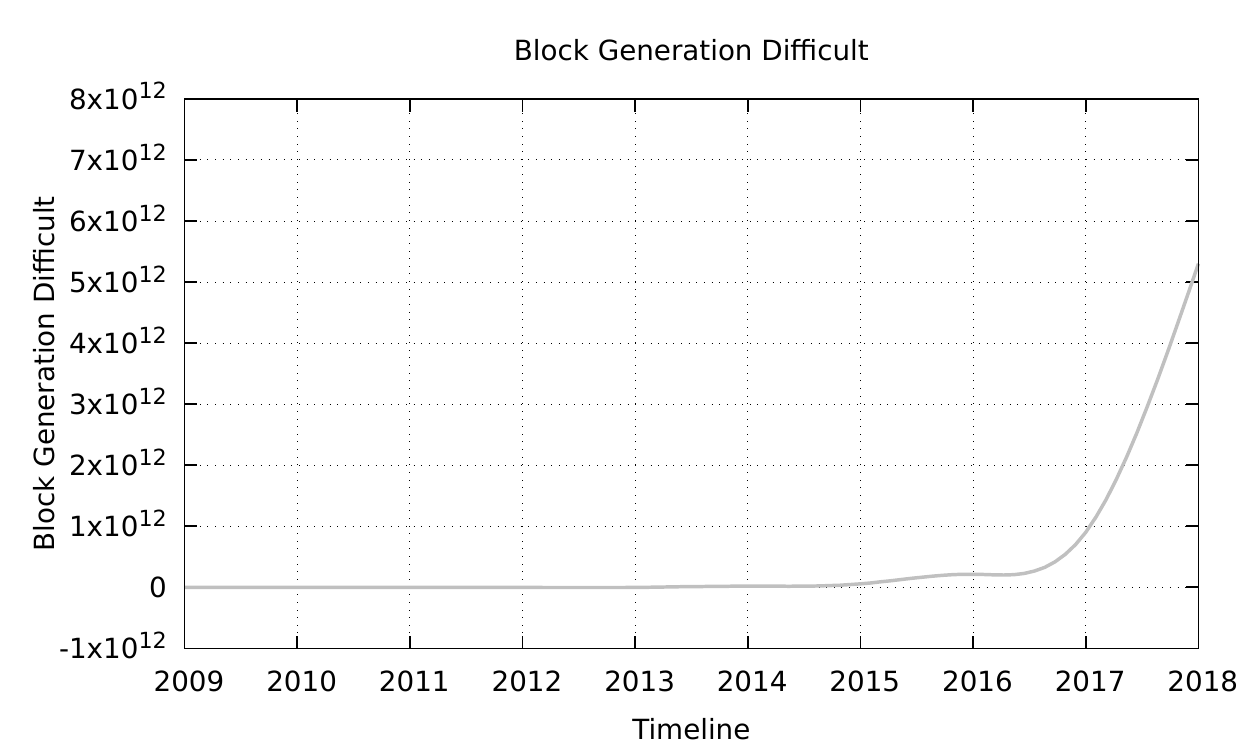}
  \caption{Difficulty to generate Bitcoin blocks \cite{databct}.}
  \label{fig:block-generation-difficult}
\end{figure}

Bitcoin's \textit{Proof-of-Work} is very similar to the problem above; where the input of the hash is given by the block herders components which include, among of other things, the timestamp, \textit{nonce} and Markle tree. The target/challenge is set by the network, considering the hash/power amount, to probabilistic generates blocks in a near-fixed time amount. \textbf{Figure \ref{fig:block-generation-difficult}} illustrates the difficulty to generate blocks in the history of the Bitcoin's Blockchain.

Although \textit{Proof-of-Work} is the most popular method to authenticate the peers in a Blockchain, there are alternatives such as the \textit{Proof-of-Stake}.

\subsubsection{Proof-of-Stake (PoS)} In Blockchain terms, Stake is what the user has and pledges in order to participate on the decision on the next block. However, unlike the name suggests the consensus is not arbitrated exclusively by the one who holds more resources, but by a set of arbitration that decides to be not centralized in one single peer. Instead, several methods have been devised.

In \textit{PoS} the miner of a new block is known as the forger. Before the selection, in order to participate in the selection party, the forger had to deposit some tokens into the network, using it as collateral to vouch for the block.

The more a user stakes, the better are the chances of being selected. An eventually malicious user will not act against the network, as it compromises the value of its tokens. Therefore, losing more money than possibly winning.

As mentioned, the key in this process is to select the right user to forge the block. This semi-random selection could be based on different factors, including Randomized block selection, Coin age-based selection, and Delegated Proof-of-Stake.

\textbf{Randomized block selection}: In the Randomized block selection, the forger is chosen based on a formula that combines the lowest hash value and the size of the stake \cite{mukhopadhyay2016brief}.

\textbf{Coin age-based selection}: A combination randomization with a "coin age" factor. The "coin age" is determinate by the age of the coin times the number of days that a coin has been held. Originally, coins that are older than 30 days are eligible to compete in the next block generation.  Older and larger sets of coins have a greater probability of signing the next block. Once the block is signed, the winning coins age is set to zero. Very old coins (older than 90 days) are no eligible to sign a block \cite{zheng2017overview}.

\textbf{Delegated Proof-of-Stake (DPoS)}: In delegated \textit{Proof-of Stake} the nodes have a reputation based on Stake. Only the 21 most reputed nodes can participate in the transaction validations and block generation. The reputation may vary upon the time; nodes can lose or conquer reputation.\cite{zheng2017overview}. 

\subsubsection{Proof-of-Authority (PoA)} transactions and blocks are validated and approved by Validators. The Validators are capable of adding data to blocks, using special software. Individuals earn the right to become a validator based on reputation. In order to keep being a validator the user most cooperate, as he/she can lose the status out of a bad reputation \cite{ProofOfAuthority}.

\subsubsection{Proof-of-Space (PoSpace)} very similar to proof of work, except that instead of computation, storage is used. Graph pebbling is one alternative for \textit{Proof-of-Space} implementation. It is a mathematical game and area of interest played on a graph with pebbles on the vertices \cite{dziembowski2015proofs}. 

\subsubsection{Ripple} Every few seconds all nodes run the Ripple consensus algorithm, maintaining correctness and agreement on the network. Once the agreement is reached the ledger is considered "closed". If there is no fork on the network, the last-closed ledger maintained by all nodes in the network will be identical \cite{baliga2017understanding}.

\subsubsection{AlgoRand} For AlgoRand decentralization means not to have to trust a centralized entity as the single source of truth in the network. The responsibility to run and maintain the network falls to ordinary users. For that AlgoRand utilizes the Byzantine Agreement: a communication protocol that allows the users of a distributed system to reach consensus in the presence of malicious actors.

\subsection{Consensus Algorithms Comparison}

All Consensus Algorithms have the same objective; however, the computation to achieve it varies a lot in between the different options. The main characteristics that were considered in this work are: \textit{node anonymity}, and \textit{power consumption}.

\begin{itemize}
  \item \textit{node anonymity}: The capability to join and exit the network without identifying itself.
  \item \textit{power consumption}: Some methods to achieve consensus demand much computation in a not so efficient way, leading to an enormous amount of energy spent.
\end{itemize}

\begin{table*}[ht]
\centering
\begin{tiny}
\begin{tabular}{ l|ccccccc }
 \hline
 \multicolumn{1}{c}{Property} & PoW & PoS & DPoS & PoA & PoSpace & Ripple & AlgoRand \\
 \hline \hline
 Anonymous &  \checkmark  & \checkmark & \checkmark & -- & \checkmark & \checkmark & \checkmark \\
 Power consumption & High & Fair & Fair & -- & Fair & Low & Low \\
 Example & Bitcoin \cite{nakamoto2008bitcoin} & Peercoin \cite{xu2016Blockchain} & Bitshares \cite{schuh2015bitshares} & -- & SpaceMint \cite{honsi2017spacemint} & Ripple \cite{pilkington201611} & AlgoRand  \\
 \hline
\end{tabular}
 \caption{\label{tab:consensusComparasion}Consensus Algorithms Comparison.}
\end{tiny}
\end{table*}

The \textbf{Table \ref{tab:consensusComparasion}} contains a comparison of the different consensus algorithms listed in this work.

\section{Software repository}
\label{chp:repositories}

Software repositories became the primary source for application installation or even application marketing. It is emerged from Linux package repositories, with later adoption by the smartphones, nowadays available in all platforms.

Many are the responsibilities/features of a package repository, including ensuring that the package is trusted, not manipulated to include any vulnerability or malware. Not less important, the repository should guarantee that there is no incompatibility among the provided packages, dependencies should be resolved upon the installation act.

Some package repositories also impose Trusted Computing Base \cite{santos2009towards} by generating signatures for each of the binaries although those are not so popular in the open source Linux repositories.

Distributed via untrusted methods such as Mirrors, the open source repositories, rely on \textit{Web-Of-Trust} in order to guarantee the integrity and authenticity of the packages. It is likely that the user will use a Mirror closer to its geographical location, therefore optimizing the download time. This mirror is not necessarily trusted or secure.

Some distributions prioritize the newest packages over old stable packages. Others also provide an alternative "repository" for a set of software or even let the user create and share their own packages via a personal repository \cite{archlinuxaur}. In that case, the user who downloads the package has to trust the package publisher, by accepting packages with her digital signature.

The timing to have the package published is sensitive, as updates may contain a security fix or even an update that will restore the machine from a failure. The timing, however, is not the only factor that is take into consideration to classify a distribution quality.

In script language repositories, where the adoption and contributions are more intensive, it is common to find orphan packages or even malicious software with names that look like valid packages \cite{ruohonen2018empirical}. Mostly as a consequence of earlier adoption. This fact reiterates the need for a figure that can bless the packages. Still, this information can be provided without centralization.

\section{Problem}
\label{chp:problem}

Having the schema of Mirrors, there is always a central repository to be mirrored from; intrinsic creates a hierarchy dependency. This repository hierarchy may delay the update of packages leading to circumstances where the user may be vulnerable to a not-yet-updated critical software failure.

The distribution from third-party repositories is not concentrated in a single information point,  not allowing the user to search for extra packages. The user has to obtain this information \textit{out-of-band}, which may or not may be trusted. It is not feasible to be indexed due to the large amount of data.

Having all packages updated to the latest version is natural. However, the majority of the distributions do not have the support to upgrade or downgrade the software up to a point in history. Having two identical platforms, rolling back in time is not a trivial task. However, it may be useful particularly in cloud computing, where virtual machine deployments are frequent. Fast code change, may lead to broken dependencies.

Concurrent trails of software may be a problem. A given package can be supplied in two or more repository sources. Likely to happens on script libraries where a package is supplied from the script repository and the distribution repository. In that case, files may clash leading to broken dependencies, broken updates and so on.

\section{Proposal}
\label{chp:proposal}

Concentrating every Linux distribution in a single package distribution source, effectively solves some of the problems explained on \textbf{section \ref{chp:problem}}, as the clash of packages or files could be identified and treated accordingly.

Having the package repository inside a Blockchain is a real benefit to the user, given the possibility of having a decentralized model for updating the packages — moreover, a centralized source for the different trails.

The proposal in this work consists in a Blockchain capable of holding packages information from different trails, this information can be later used in package upgrades and downgrades, search and installation.

We present a Blockchain that does not have financial incentives to the participants. The benefit, on the best interest of the distributions, is the publication of the packages. The packages are later consumed by the distribution users, for every confirmed download a reward is granted to the associated distribution. The interest to have a trustworthy source for packages is mutual and interdependent.

Ultimately, the distributions with largest users base are the ones most interested to keep the quality of the Blockchain, therefore trusted to forge new blocks. The popularity factor is used to measure how large is a distribution user base. The popularity factor is computed based on the reward given for the distribution at every confirmed package download.

The concept presented in this work is somewhat similar to the concept of \textit{Proof-of-Stake}, where the Stake factor is the popularity given the amount of downloads related to each distribution. Inside the Blockchain the distribution will be called as Trail.



\subsection{Trail}

The Trail is the representation of a distribution. Multiple users can act on behalf of a Trail. Within the Blockchain, it is possible to add, list, remove users and packages from a given Trail. Similar to the packages information, the user authorization is also hosted within the Blockchain. There are no permissions levels detailed in this work, although it may be expanded to support it in further versions.

An authorization is a grant to a public key. Whoever had the associated private key will be able to manipulate the Trail, including the removal of its participation on the Trail. New trails can be created; however, the creation depends upon a process that is held on multiple Blocks of the Blockchain as detailed on {\bf Creating a new Trail}.

\subsubsection{Creating a new Trail}

Any member of the Blockchain can generate a new Trail, no authority is needed to validate it or impose restrictions, although it will not be possible to request a name that is already taken by another user. The ownership of a name is only released via hard-fork, or if there isn't a member to claim for the name. A block that offers to a user a trail that is already taken, is considered to be an invalid block; regardless of who offered it.

In order to avoid the massive generation of new Trails, each block can only contain ten new Trails. The preference is given to the users that are already part of a popular Trail (The concept of popularity will be explained on subsection \ref{trail_popularity_proof-of-download}). The generation is also taken in two different blocks: {\bf request} and {\bf response}. In the request phase, the user asks the network for the Trail, presenting her public key. If eligible to be in the next block, the next block will contain a challenge that is encrypted and can only be read by whoever has the corresponding private key. The second step is to provide this result in the peer-to-peer network, in case the challenge is resolved, the Trail will be published in the second block.

\begin{figure}[!ht]
  \centering
  \includegraphics[scale=0.40]{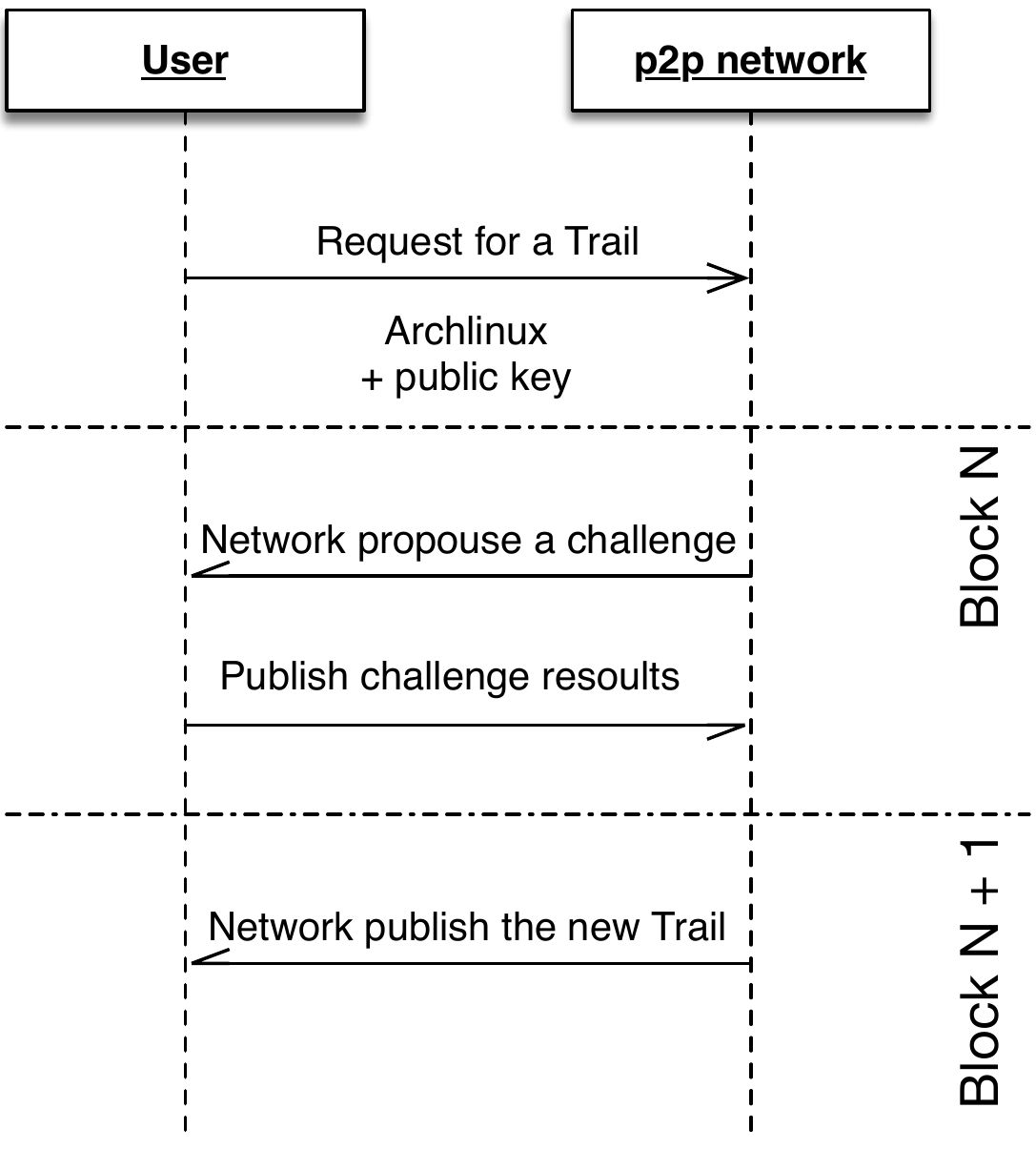}
  \caption{Diagram that illustrates the creation of a new Trail.}
  \label{fig:new-trail-creation}
\end{figure}

If everything proceeds with the regular flow (no exceptions), the Trail is formalized, and at a third block, it will be possible to add more identities to the Trail. The \textbf{Figure \ref{fig:new-trail-creation}} contains a flow that describes the creation of a new Trail.

\subsubsection{Deleting a Trail}

The Trail is deleted whenever no more users are holding the ownership. If that happens, the trail name will be vacant for the next user who claims it. Notice that the packages blessed to a given trail are still valid, no matter if valid users are holding the Trail authority or not.

\subsubsection{Adding an user to a Trail}

Adding a user to a trail is conditioned to:

\begin{itemize}
    \item Someone who already have privileges on the given trail grants access to a third user.
    \item The third user in question accepts the invitation by solving a cryptographic puzzle.
\end{itemize}

\begin{figure}[!ht]
  \centering
  \includegraphics[scale=0.35]{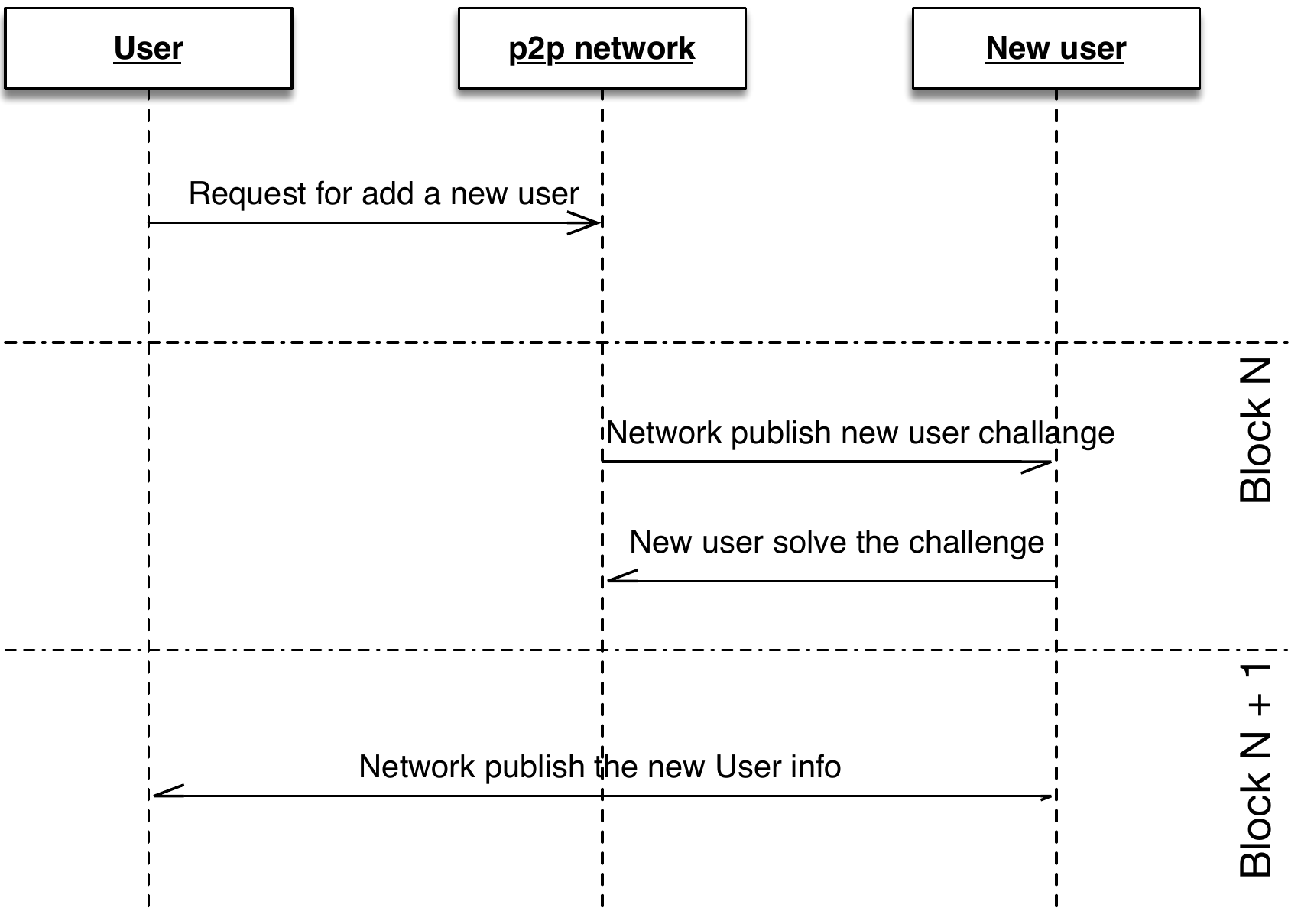}
  \caption{Sequence diagram that illustrates the addition of an user into a Trail.}
  \label{fig:adding-user-to-trail}
\end{figure}

With these simple rules, it is possible to guarantee that Trails will not be held by fake users (invalid), furthermore will not be taken by someone that is not interested in having such access. The puzzle is a challenge-response problem that depends on the user to use its private key. When added to a Trail, the user is known as Trail member. \textbf{Figure \ref{fig:adding-user-to-trail}} illustrates the sequence flow that describes the addition of an user to a Trail.

\subsubsection{Removing an user from a Trail}

Differently from adding a user to a Trail, to remove the user, there is no puzzle. Instead, another member of the Trail, or the user itself, can publish the removal information on the Blockchain.

\begin{figure}[!ht]
  \centering
  \includegraphics[scale=0.35]{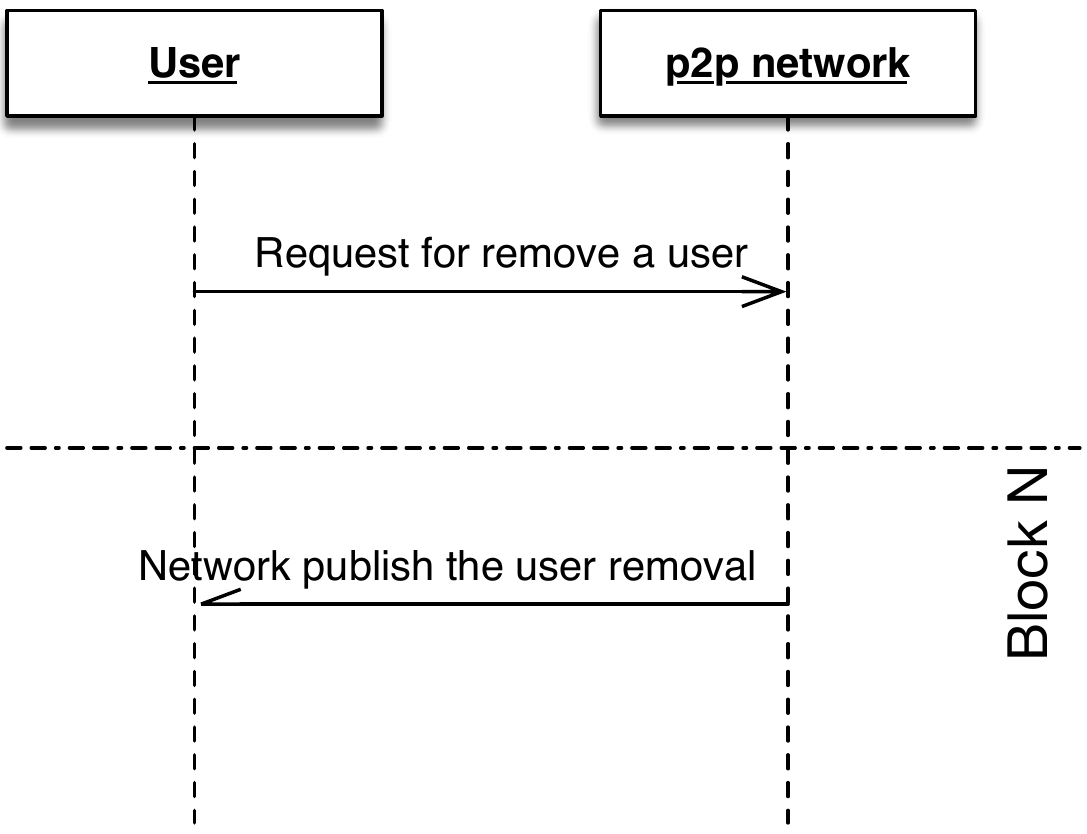}
  \caption{Sequence diagram that illustrates the removal of a user from a Trail.}
  \label{fig:remove-user-from-trail}
\end{figure}

Notice that it is technically possible to have a nested Trail. That may happen when the Trail is held only by users that no longer have access to their certificates. If that happens, the user won't be able to remove himself from the Trail.
\textbf{Figure \ref{fig:remove-user-from-trail}} illustrates the sequence flow that describes the removal of a user from a given Trail.

\subsubsection{Trail, distributions and version names}

There are Linux and packages distributions that do not have a version associated with it. That is the case of ArchLinux, and PiPy; other distributions such as Debian contain different releases with different packages set for each version. In this second case, it is recommended that the distribution have a Trail for each version.

\subsubsection{Adding packages to a Trail}

\begin{figure}[!ht]
  \centering
  \includegraphics[scale=0.35]{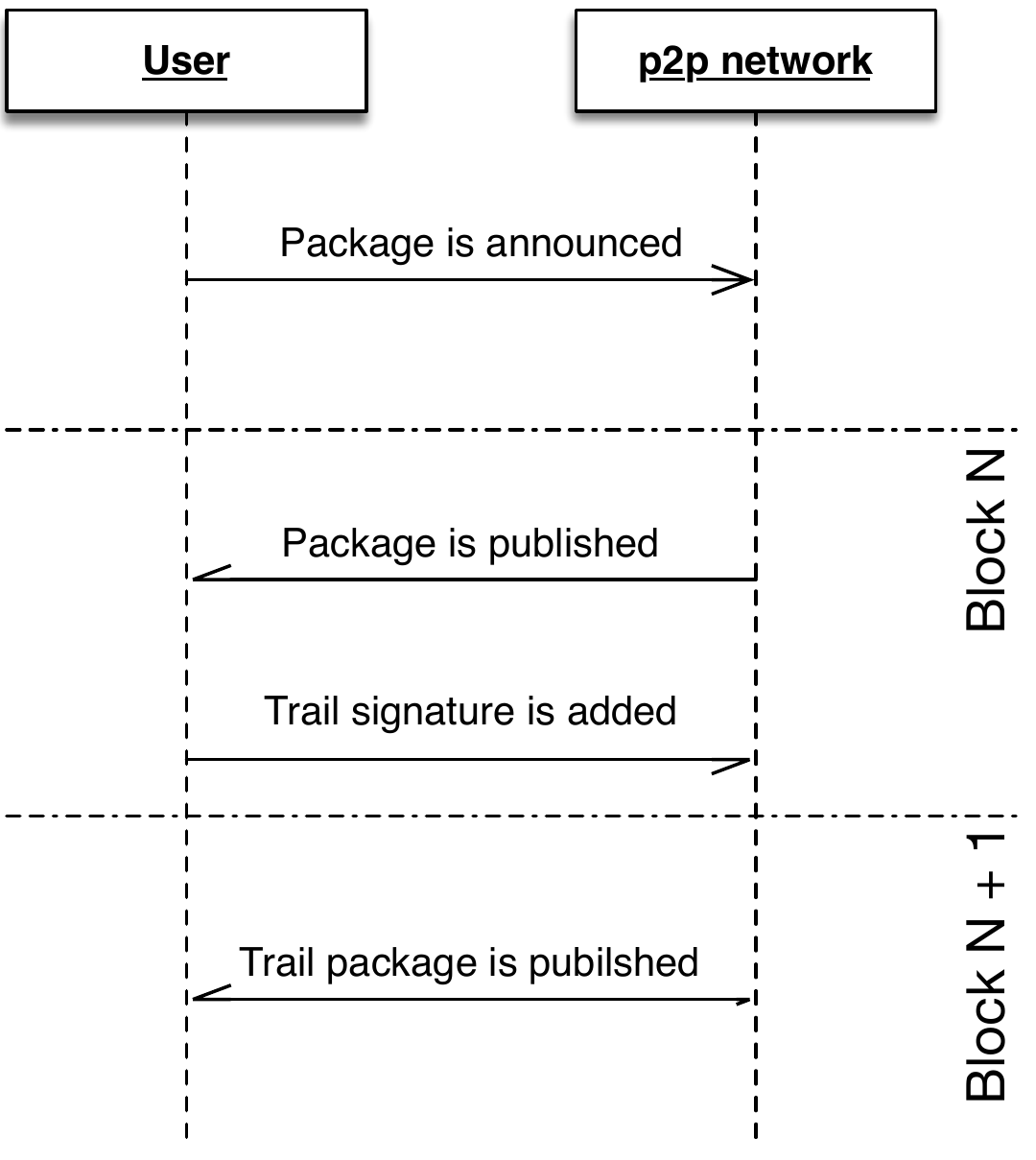}
  \caption{Diagram that illustrates the addition of a package to the Blockchain.}
  \label{fig:adding-package-to-trail}
\end{figure}

There is no such thing as adding a package to a Trail. The package is added to the Blockchain, and when added, a Trail member can vouch for it. The vouch is a simple signature provided by any Trail member. The signature consists in the package check sum and the Trail name. That will guarantee that a given Trail trusts in the package as shown on \textbf{Figure \ref{fig:adding-package-to-trail}}.

There is a limit of 100 packages to be published per block, the equation gives the preference on how popular is the Trails that the user who is publishing the package is a member. \textbf{Equation \ref{eq:popularity_user_pak_publish}}. The Preference is given as the sum of the popularity of each Trail that the user is member.

\begin{equation}\centering\label{eq:popularity_user_pak_publish}
P = 
\sum_{n=0}^{t} t
\end{equation}

\subsection{The block generation: consensus}
\label{the_block_generation_consensus}
Although a single entity will be responsible for generating a block, any member of the Blockchain can validate the Block. A set of simple rules can be used to determine if a block is valid or not, as listed below:

\begin{itemize}
    \item The Block respects the limits on the number of new Trails and Packages.
    \item The Signature of the Trail members are valid, and the publishers are indeed authorized to vouch for that Trail.
    \item The new Trail members request and removal match the signatures.
    \item The new Trail members have correctly solved the proposed puzzle.
\end{itemize}

All trail members are eligible to forge a Block. During the forging process, a group of the most four popular Trails is selected. Each Trail is represented by a randomly chosen member. One, out of those four, will be randomly chosen to forge the Block.

\subsection{Trail popularity: proof-of-download}
\label{trail_popularity_proof-of-download}

The proof-of-download is utilized to ensure the popularity of a Trail. It consists in a process to make sure that a client that is downloading a package is a legitimate client, not an adversary trying to increase the popularity of a given trail; in order to achieve this goal, there is a challenge posted to the client. The challenge itself inputs a minimum overhead to the client and an even smaller overhead to the provider.

The client has to:

\begin{itemize}
    \item Inform the package name and trail;
    \item Download the package and challenge;
    \item Compute the hash of the downloaded package;
    \item Give back the hash to the p2p network with challenge receipt;
    \item If valid, the network will give back the offset of chunk data that needs to be removed from the downloaded package; In that case the download count is increased. Anyone is able to validate the challenge. If the challenge was already solved by a previous client, it won't be summed. 
    \item If not valid, it will be ignored.
\end{itemize}

Then every peer on the network, when receiving a request to deliver a package has to:

\begin{itemize}
    \item Open the file and tamper the package with random data.
    \item Compute the hash of the package.
    \item Delivery the package with the tampered content.
    \item Whenever it gets the request to validate a hash, validate and return the data.
    \item Increment the ledger within the download numbers.
\end{itemize}

This number is volatile and held by the network; a snapshot of this number is published on every new Block. So, every node is fully aware of the most popular Trails.

The popularity rate is directly proportional to the amount of downloads.
Every block computes the popularity considering: current download rate
and previous popularity. As expressed in the equation: \textbf{Equation \ref{eq:popularity_rate}}. \textbf{P} stands for popularity while \textbf{pp} is previous popularity, \textbf{cp} is currently popularity, and \textbf{t} is total amount of downloads.

\begin{equation}\centering\label{eq:popularity_rate}
P = ((pp/100)*t) * 0.3) + (cp * 0.7)
\end{equation}

The number of downloads needs to be well computed as it may grant access to the generation of the new Block. With that in mind, the network imposes a challenge to every new download. The challenge consists of tampering the original download file with a random set of bytes. Once the tamper is made, the HASH of the package is computed.

Once downloaded, the package needs to be verified by the client, once the client got the hash, it gives it back to the network, which only presents the tamper formula, if the hash is presented correctly.

In this scenario, fake clients need to download the package making it more expensive than the challenge generation, therefore mitigating the possibility of having fake downloads only to inflate those numbers. The cost of the challenge is inexpensive as the package needs integrity validation nevertheless.

\section{Experiment}
\label{chp:experiment}

The experiment consists in a creation of a test Blockchain, where all packages from the ArchLinux \cite{archlinux} distribution were placed, so that, it was possible to understand the feasibility of haven a distribution in the proposed format. With the experimental Blockchain, it is also possible to analyze factors such as block size, perfect timing for block generation, and format and disposition of the data for the different \textbf{Trails}. Most importantly, the experiment validates  the rules proposed on section \ref{the_block_generation_consensus}.

During the experiment, the signature creation was laid aside, since they have been proved to work elsewhere \cite{pointcheval2000security}. Without the signature generation, it was possible to drastically increase the simulation speed.

ArchLinux was chosen because it has the right balance on stability and package update frequency. It also has a 3rd party repository AUR \cite{archlinuxaur} which will be handled in a further work.

The simulation deals with almost 5000 users publishing packages, over ten years of ArchLinux package history. The simulation assumes the distribution popularity based on educated random numbers.

During the simulation limits on the number of packages per block are reached, and a consensus is eventually met.

The simulation software does not test the p2p network; rather, it tests the manner in which to achieve consensus. Notice, however, that within the block interval the package creation time is irrelevant, as the acknowledgment of its existence to every peer. Therefore this experiment also proves (to some degree) that the suggestions here are resilient to problems arising from the network, be it intentionally caused by an eventual attacker or not.

The presence of a bad actor trying to subvert the consensus does not need to be expressed, as every peer has its interest and advocate on its behalf, yet, by the proposed rules a common consensus is meant to be met.

All implementations created and used in this work are available on GitHub licensed under GPL \cite{githubit}. 

\subsection{The Test Blockchain}

In ArchLinux the package creation recipes are placed in a Git \cite{git} repository \cite{archpackages}. The package receipt contains all the information that is necessary to validate the packages sources and compile them. This receipt format contains variables and metadata that was interpreted in order to produce more consistent results. In the \textbf{Figure \ref{fig:pkgbuild}} there is an example of a PKGBUILD file. Notice the hash for every relevant source in the package.

\begin{figure}[!ht]
\centering
\begin{tiny}
\centering
\begin{verbatim}
pkgname=openssh
pkgver=7.9p1
pkgrel=1
pkgdesc='Premier connectivity tool for remote login with the SSH protocol'
url='https://www.openssh.com/portable.html'
license=('custom:BSD')
arch=('x86_64')
makedepends=('linux-headers')
depends=('krb5' 'openssl' 'libedit' 'ldns')
optdepends=('xorg-xauth: X11 forwarding'
            'x11-ssh-askpass: input passphrase in X')
validpgpkeys=('59C2118ED206D927E667EBE3D3E5F56B6D920D30')
source=("https://ftp.openbsd.org/pub/OpenBSD/OpenSSH/portable/
            ${pkgname}-${pkgver}.tar.gz"{,.asc}
        'sshdgenkeys.service'
        'sshd@.service'
        'sshd.service'
        'sshd.socket'
        'sshd.conf'
        'sshd.pam')
sha256sums=('6b4b3ba2253d84ed3771c8050728d597c91cfce898713beb7b64a305b6f11aad'
            'SKIP'
            '4031577db6416fcbaacf8a26a024ecd3939e5c10fe6a86ee3f0eea5093d533b7'
            '3a0845737207f4eda221c9c9fb64e766ade9684562d8ba4f705f7ae6826886e5'
            'c5ed9fa629f8f8dbf3bae4edbad4441c36df535088553fe82695c52d7bde30aa'
            'de14363e9d4ed92848e524036d9e6b57b2d35cc77d377b7247c38111d2a3defd'
            '4effac1186cc62617f44385415103021f72f674f8b8e26447fc1139c670090f6'
            '64576021515c0a98b0aaf0a0ae02e0f5ebe8ee525b1e647ab68f369f81ecd846')
\end{verbatim}
\end{tiny}
  \caption{Relevant parts of the PKGBUILD file for the OpenSSH package.}
  \label{fig:pkgbuild}
\end{figure}

As this package receipt format was updated over time, our work also includes a given version to the parser, in order to make this experiment reproducible.

The creation of the test Blockchain work is divided into two different pieces: \textit{Capturing and processing packages information} and \textit{Constructing the Blockchain}.

\subsubsection{Capturing and processing packages information:} ArchLinux packages Git repository contains all the changes in every package of the distribution. For every change, there is a new commit. A single commit may also hold modification for different packages. Within the package recipe, there are, among other things, the package name and version. There is no package index; instead, there is the repository history.

By enumerating all the commits from the package repository, it was possible to parse the changes and identify the history of each package for later inclusion in a Blockchain. In order to do that, a Python script was created to iterate over all commits grabbing the differences. For every new package change, the package recipe was parsed, and the 'version' was saved, indexed by the change date.

The process to extract data from the Git repository is I/O intensive, in order to process the data within a reasonable amount of time a ramdisk\footnote{ramdisk: File system created atop of the RAM. Reducing data seek time and increases the throughput.} was used. The process was not prompt, therefore splitting into two stages. Once the data was sorted by date, the process to construct a Blockchain was started.

\subsubsection{Constructing the Blockchain:} In this experiment, we are simulating a Blockchain in which consensus will be achieved among users that were added as part of the simulation as well. The blocks generation will assume the block timestamp a date in the past, respecting the time when the packages were created.

Four valid Trails were added to the simulation: \textbf{archlinux}, \textbf{pypy}, \textbf{perl}, \textbf{ruby}, other Trails were also added. The full list of the Trails used on the simulation is available at \textbf{Table \ref{tab:trailregex}}. To decide which package goes to which Trail, there is a regular expression for each Trail. If the regular expression matches the package name, the package is considered to be part of the Trail. Notice that a package can be vouched for more than one Trail. 

\begin{scriptsize}
\begin{table}
\centering
\begin{tabular}{p{0.05cm} p{10cm}}
(a) & \begin{tabular}{ c c c c }
 \hline
 Trail ( Regex ) \\
 \hline
 \hline
archlinux (.*) & perl (perl) & pypy (py) & ruby (rb) \\ 
 \hline
\end{tabular}
\end{tabular}
\newline
\newline
\newline
\begin{tabular}{p{0.05cm} p{10cm}}
(b) & \begin{tabular}{c c c c}
 \hline
 Trails \\
 \hline
 \hline
ALTLinux & Ark Linux & BasicLinux & BioKnoppix \\
CentOS & Conectiva Cucumber Linux & Debian & Zenwalk Linux \\ 
Devil-Linux & Dyne:Bolic & Feather & Floppix   \\ 
Freesco &  Frugalware &  Gentoo &  Gnoppix   \\ 
IPCop & Kanotix & Knoppix & Kurumin   \\ 
Linux Scratch &  Lycoris &  Manjaro & Morphix   \\ 
Pardus &  PHLAK & Puppy Linux & Red Hat Ent  \\ 
SLAX & Source Mage &  SuSE & TopologiLinux   \\ 
Turkix & Univention Corp & Whitebox Linux & Yoper \\
Amigo Linux & BackTrack & BeatrIX & BLAG  \\ 
ClusterKnoppix & CRUX & DamnSmallLinux & DeLi Linux   \\ 
DragOnLinux & Elive & Fedora & Foresight  \\ 
Freespire & G2Linx & Goodgoat & GoboLinux  \\ 
IpodLinux & Kate OS & Kubuntu &  Linspire  \\ 
Lunar Linux & Mandriva & MEPIS & muLinux   \\ 
PCLinuxOS & PocketLinux & Red Hat & Slackware   \\ 
SmoothWall & Sun JDS & SystemRescue & TurboLinux  \\ 
Ubuntu Linux & VectorLinux & Yellow Dog & Xandros \\
\hline
\end{tabular}
\end{tabular}

 \caption{\label{tab:trailregex}Regular expression used to check whether or not a package belongs to a given trail. (a) Trails with custom regular expressions. (b) Trails where the regular expression consists in the first four digits of the Trail name.}
\end{table}
\end{scriptsize}

The vouch action is, naturally, placed after the addition of each package. During the simulation, the vouch was programmed to happen from the next block to four blocks ahead; The decision as to when to vouch for respects the probability stated on \textbf{Table \ref{tab:vouchprob}} 

\begin{table}
\centering
\begin{tabular}{ c|c }
 \hline
 Block number & Probability\\
 \hline \hline
 N + 1 & 60\% \\ 
 N + 2 & 20\% \\  
 N + 3 & 10\% \\
 N + 4 & 10\% \\
 \hline
\end{tabular}
 \caption{\label{tab:vouchprob}Probability when to vouch for a given package.}
\end{table}

The genesis block was generated using a timestamp 40 minutes before the first data in the git repository, allowing the Trail ArchLinux to be created. The interval between the block generation was chosen to be 20 minutes. A change in this value can affect the size of the blocks.

Atop of the regular data that is part of the block, the simulation also adds some metadata information, useful to understand the way in which the consensus was achieved. \textbf{Figure \ref{fig:block}} contains an example of a block.

\begin{figure}[!ht]
\centering
\begin{tiny}
\begin{verbatim}
{
 "forger": "Poppy",
 "metadata": {
  "amount_of_packages": 1, "amount_of_valid_trails": 4, "everybody_that_can_forge_this_block": [
   { "popularity": 38.60318029211056, "trails": [ "pypy" ], "user": "Ava" },
   { "popularity": 36.28520425873052, "trails": [ "archlinux" ], "user": "Poppy" },
   { "popularity": 13.26900777134668, "trails": [ "perl" ], "user": "Leo" },
   { "popularity": 11.842607677812248, "trails": [ "fal" ], "user": "Amelia" }  ],
  "popularity_at_generation": [
   { "name": "fal", "pop": 5.2051448090322765 },
   { "name": "archlinux", "pop": 39.20033948972628 },
   { "name": "perl", "pop": 15.810333717513881 },
   { "name": "pypy", "pop": 39.78418198372756 }  ]  },
 "number": 150,
 "packages": [ {
   "package": {
    "arch": [ "i686", "x86_64" ], "depends": "python", "license": [ "GPL" ],
    "md5sums": "1af233c6fa0a68851bc6155b2f563c30", "name": "bzr",
    "parser": "regexp v1.0", "pkgrel": 1, "pkgver": "1.3",
    "pkgdesc": [ "A decentralized revision control system" ],
    "source": "http://bazaar-vcs.org/releases/src/bzr-$pkgver.tar.gz",
    "url": "http://www.bazaar-vcs.org"  },
   "publisher": { "name": "Pammi", "signature": "signature goes here" }  } ],
 "popularity": [
  { "name": "fal", "pop": 11.842607677812248 },
  { "name": "archlinux", "pop": 36.28520425873052 },
  { "name": "perl", "pop": 13.26900777134668 },
  { "name": "pypy", "pop": 38.60318029211056 }  ],
 "trails": []
}
\end{verbatim}
\end{tiny}
  \caption{Example of a Blockchain block inside the simulator}
  \label{fig:block}
\end{figure}

The educated values assumed for download numbers were expressed on \textbf{Table \ref{tab:downloadnumbers}}. Those numbers attempt to mimic the default behavior of a live system. On the next section, it is explained how those numbers were computed.

\begin{table}
\centering
\begin{tabular}{ l|c|c }
 \hline
 Trail & Minimum value & Maximum value \\
 \hline \hline
 archlinux & 200000 & 300000 \\
 pypy & 10000 & 400000 \\
 perl & 100000 & 100500 \\
 ruby & 100000 & 100900 \\
 others & 0 & 100000 \\
 \hline
\end{tabular}
 \caption{\label{tab:downloadnumbers}Range utilized to simulate the number of confirmed downloads.}
\end{table}


\section{Results}
\label{chp:results}



\begin{figure}[!ht]
  \centering
  \includegraphics[scale=0.65]{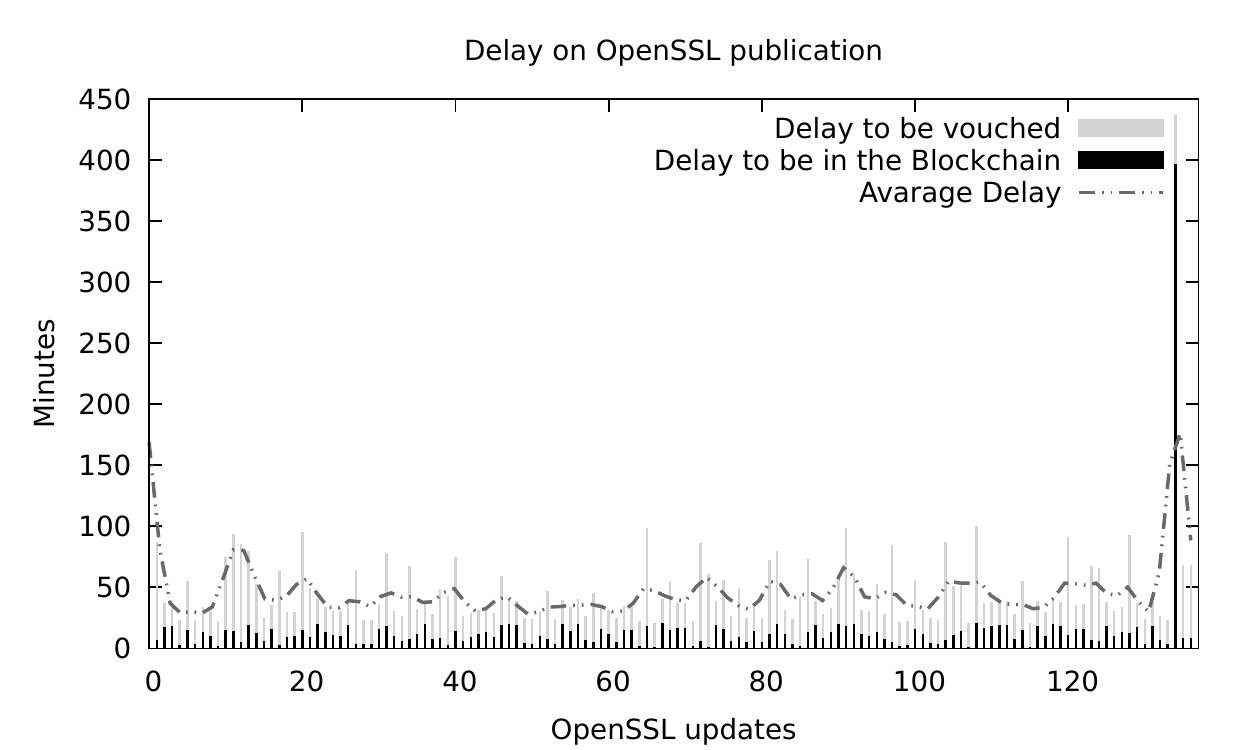}
  \caption{Delay to have the OpenSSL package available to the Trail users.}
  \label{fig:openssl_pub}
\end{figure}

The simulation of the Blockchain was fundamental to validate the proposal of this work. Within the simulator, it is possible to validate the score on the distribution popularity, and also to verify that all packages for ArchLinux were there. Some of the packages have a significant delay to be published, as expected due to block limitation on the number of packages. \textbf{Figure \ref{fig:openssl_pub}} shows the delay on the OpenSSL publication.

\begin{figure}[!ht]
  \centering
  \includegraphics[scale=0.65]{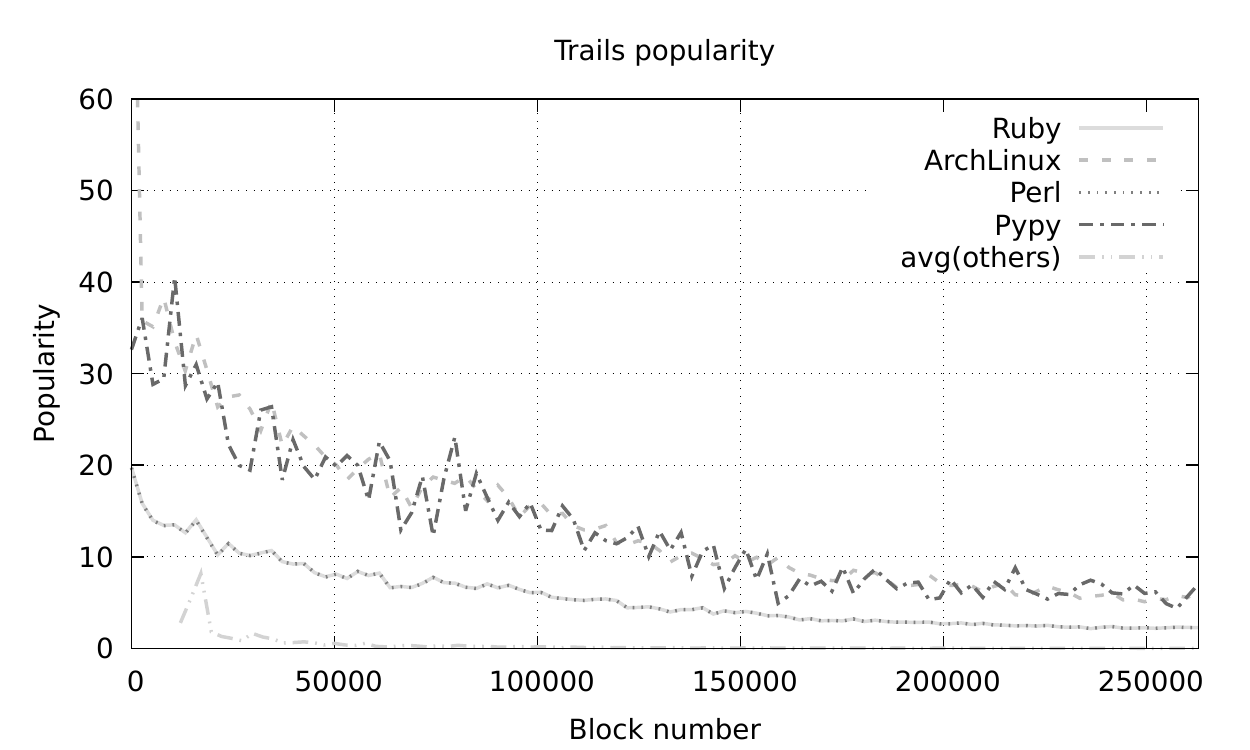}
  \caption{Popularity of the trails along the block generation.}
  \label{fig:trails_reputation}
\end{figure}

On \textbf{Figure \ref{fig:trails_reputation}}. It is possible to find different trials' reputation during the blocks generation. The reputation is consistent, in a sense that there is no chance of unpopular Trails generating the Blocks, therefore limiting a malicious user to create a new Trail, to somehow tamper the forgery of a package.

\begin{figure}[!ht]
  \centering
  \includegraphics[scale=0.45]{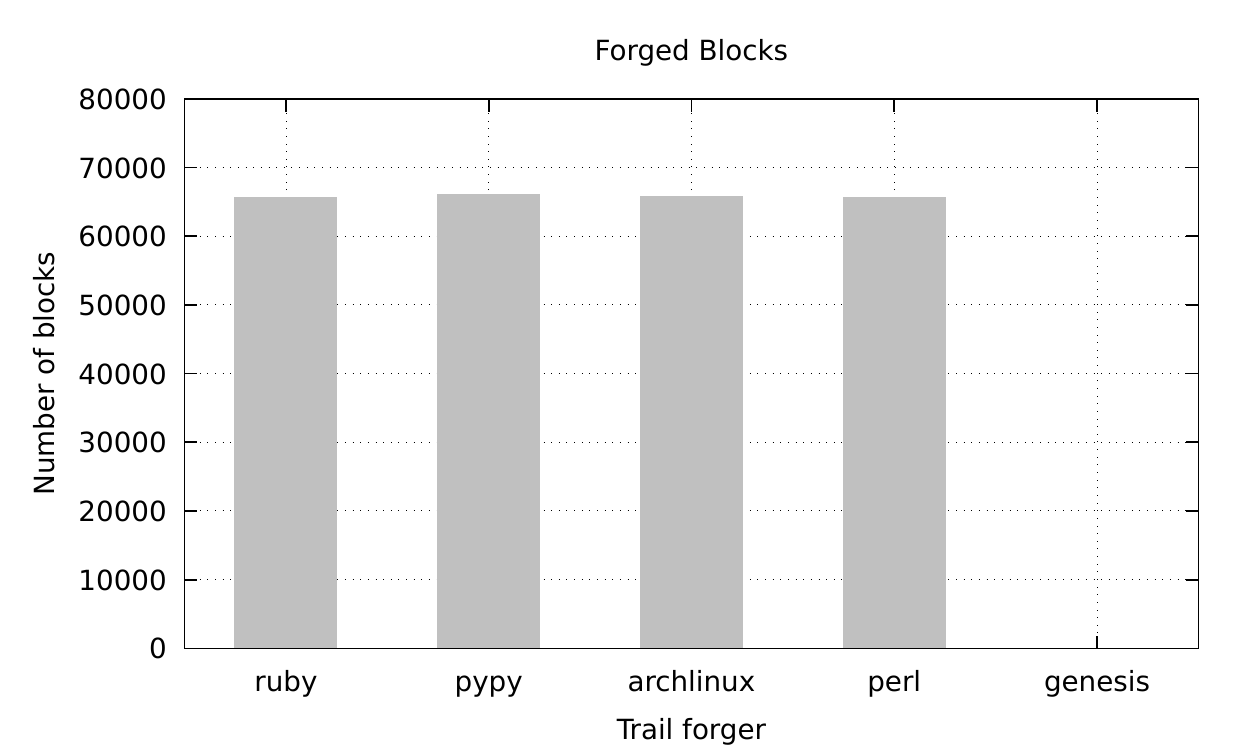}
  \caption{Trails that forged Blocks during the experiment.}
  \label{fig:block_forgers}
\end{figure}

It is also possible to notice that the distribution with less popularity did not manage to forge any package; This shows that it is indefeasible to create new Trails to tamper the Blockchain. \textbf{Figure \ref{fig:block_forgers}} summarizes all forge blocks.

\begin{figure}[!ht]
  \centering
  \includegraphics[scale=0.45]{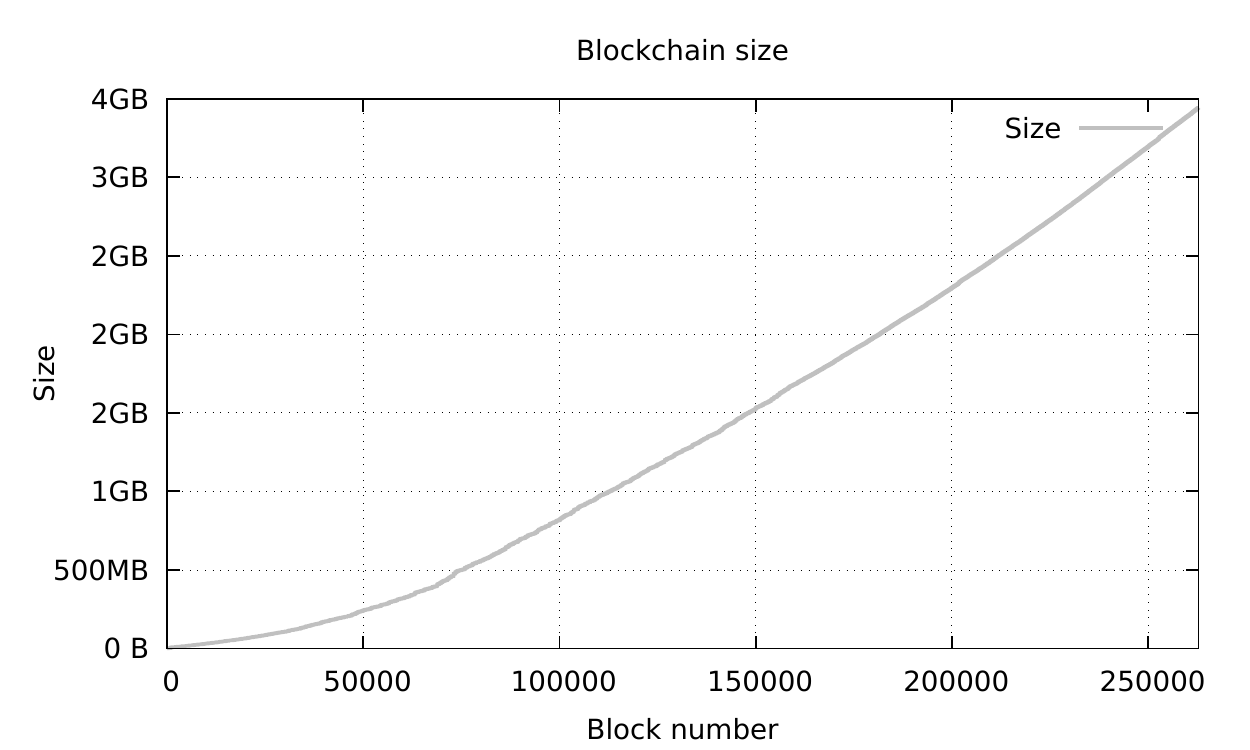}
  \caption{Blockchain size by block.}
  \label{fig:blockchain_size}
\end{figure}

The size of the Blockchain structure was not expressive, given the amount of data stored. For production, the JSON format could be replaced with a more optimal data. It is still interesting to keep JSON in the simulation for the easy readability. \textbf{Figure \ref{fig:blockchain_size}} illustrates the size of the Blockchain increasing by each block.

\begin{figure}[!ht]
  \centering
  \includegraphics[scale=0.45]{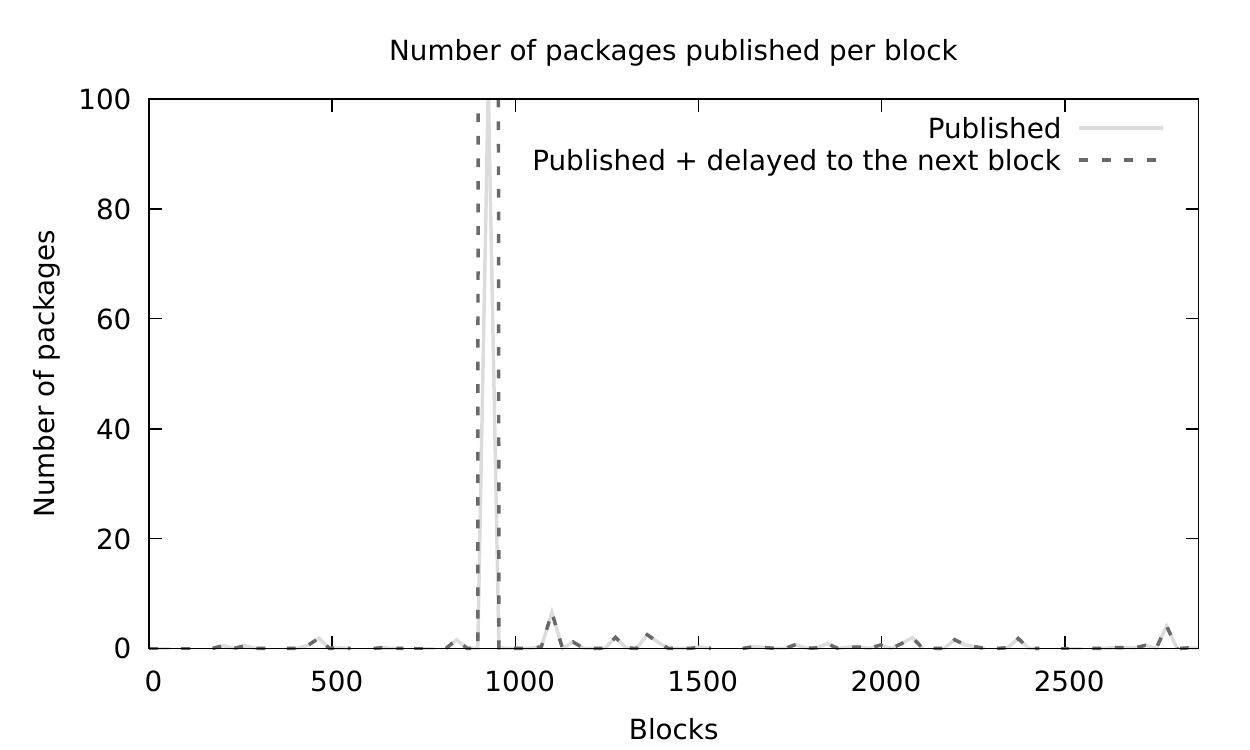}
  \caption{Number of packages published per block.}
  \label{fig:num_per_block}
\end{figure}

\begin{figure}[!ht]
  \centering
  \includegraphics[scale=0.45]{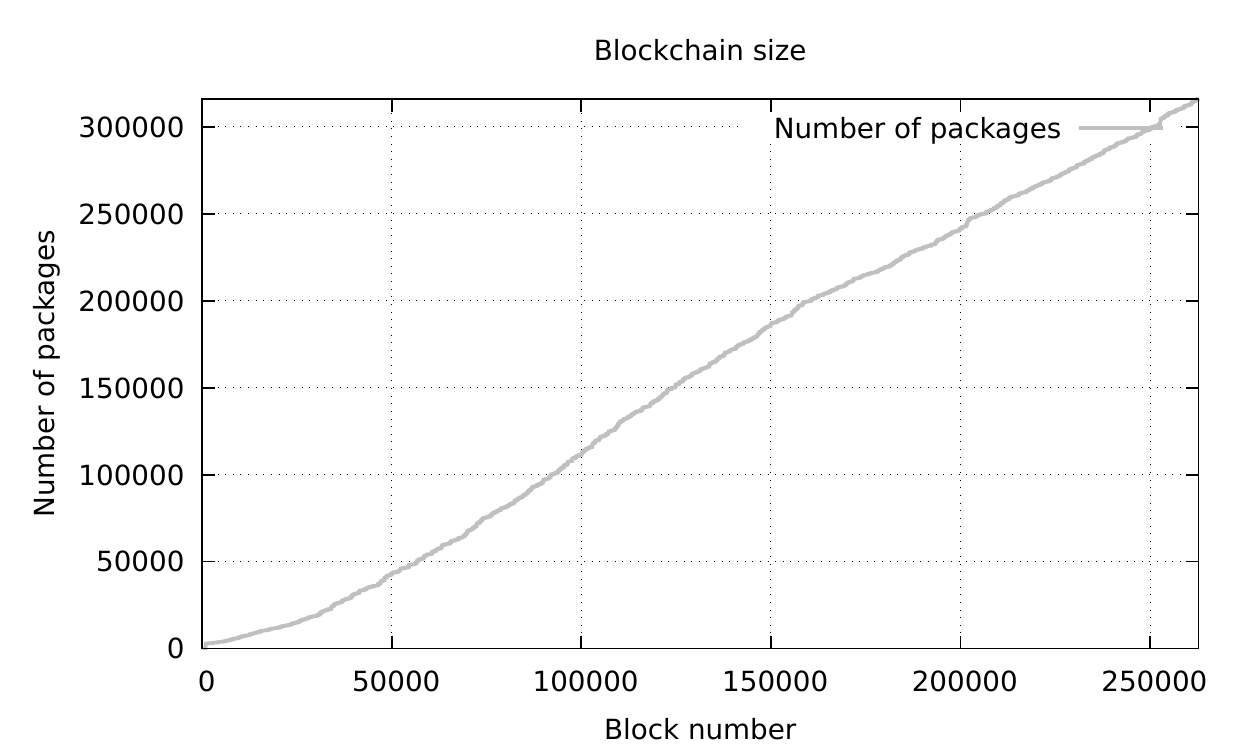}
  \caption{Number of packages in the Blockchain.}
  \label{fig:num_per_blockchain}
\end{figure}

At Block 923 (\textbf{Figure \ref{fig:num_per_block}}) one can observe the first time that the number of published packages hits the block limit. Important to notice that most popular users (given the fact that they are members for popular Trails) got their packages published first. The others got the packages published on best effort on the upcoming blocks. \textbf{Figure \ref{fig:num_per_blockchain}} presents the amount of packages on the Blockchain.

\subsection{Security Analysis}
The Blockchain proposed in this work guarantees that the user will download the package vouched for the distribution that he/she chose to trust, an adversary will not be able to poison the Blockchain with "trusted" packages that could contain malicious software. To ensure that, the Blockchain has the following features:

(a) All published packages are digitally signed.
(b) All the vouch actions are digitally signed, ensuring: Authentication, Integrity, and Non-repudiation.
(c) The addition or removal of users as members of the Trails are based on digital signatures as well.
(d) The block is also signed with the forgery digital signature.
(e) The forgery is chosen upon popularity (as demonstrated in the experiment). Therefore, keeping the most interested in the correctness of the Blockchain as the responsible, to generate the next Block.
(f) Popularity is granted to verified downloads only.

\subsubsection{Creating the most popular Trail:} Technically it is feasible to create a new Trail that will be more popular than the authentic ones. However, to do that, it will be necessary to have a more significant amount of downloads than the authentic one, for quite some time. Thus, the effort to tamper the Block generation increases proportional to the importance of the authentic Trails, given its popularity. 

\subsubsection{Certificate hijacking:} It was not yet proposed a revoke mechanism. In case of a digital certificate hijack, the attacker will be able to impersonate a Trail member.

\section{Conclusion}
\label{chp:conclusion}

Through simulation we have tried to demonstrate that the Blockchain consensus was met as expected by the Blockchain members, together with the prioritization of the most important packages.

The overhead given by the Blockchain utilization was not significant, given the amount of data. There was however an average delay of 39 minutes on the publication of the packages to the end user. That delay was a consequence of the Block generation and the posterior Vouch action.

The delay in the publication is acceptable as, normally the packages need to be transmitted to mirrors which is likely to have the same or even bigger delays.

The proof-of-download was proved to be efficient as effectively confirm the download of the package, therefore producing consistent numbers on the popularity of the Trails.


\bibliography{bibliography}

\begin{thebibliography}{10}

\bibitem{andreas2014mastering}
Andreas Antonopoulos.
\newblock {\em Mastering Bitcoin}.
\newblock O{'}Reilly, Media, 2 edition, 5 2017.

\bibitem{archlinuxaur}
aurweb Development~Team.
\newblock Aur home.
\newblock \url{https://aur.archlinux.org/}, 2019.
\newblock Accessed: 2019-01-13.

\bibitem{baliga2017understanding}
Arati Baliga.
\newblock Understanding blockchain consensus models.
\newblock {\em Persistent}, 2017.

\bibitem{caronni2000walking}
Germano Caronni.
\newblock Walking the web of trust.
\newblock In {\em Enabling Technologies: Infrastructure for Collaborative
  Enterprises, 2000.(WET ICE 2000). Proeedings. IEEE 9th International
  Workshops on}, pages 153--158. IEEE, 2000.

\bibitem{databct}
Kacper Cieśla.
\newblock Average time to mine a block in minutes.
\newblock \url{https://data.bitcoinity.org/bitcoin/block_time/all?t=l}, 2018.
\newblock Accessed: 2019-01-05.

\bibitem{git}
Software~Freedom Conservancy.
\newblock Git is a free and open source distributed version control system.
\newblock \url{https://git-scm.com/}, 2019.
\newblock Accessed: 2019-01-13.

\bibitem{dziembowski2015proofs}
Stefan Dziembowski, Sebastian Faust, Vladimir Kolmogorov, and Krzysztof
  Pietrzak.
\newblock Proofs of space.
\newblock In {\em Annual Cryptology Conference}, pages 585--605. Springer,
  2015.

\bibitem{honsi2017spacemint}
Trond H{\o}nsi.
\newblock Spacemint-a cryptocurrency based on proofs of space.
\newblock Master's thesis, NTNU, 2017.

\bibitem{BlockChainTruth}
Marco Iansiti and Karim~R. Lakhani.
\newblock The truth about blockchain.
\newblock \url{https://hbr.org/2017/01/the-truth-about-blockchain}, 2017.
\newblock Accessed: 2019-01-04.

\bibitem{mukhopadhyay2016brief}
Ujan Mukhopadhyay, Anthony Skjellum, Oluwakemi Hambolu, Jon Oakley, Lu~Yu, and
  Richard Brooks.
\newblock A brief survey of cryptocurrency systems.
\newblock In {\em Privacy, Security and Trust (PST), 2016 14th Annual
  Conference on}, pages 745--752. IEEE, 2016.

\bibitem{nakamoto2008bitcoin}
Satoshi Nakamoto.
\newblock Bitcoin: A peer-to-peer electronic cash system, 2008.

\bibitem{okupski2014bitcoin}
Krzysztof Okupski.
\newblock Bitcoin developer reference.
\newblock {\em Eindhoven}, 2014.

\bibitem{pilkington201611}
Marc Pilkington.
\newblock 11 blockchain technology: principles and applications.
\newblock {\em Research handbook on digital transformations}, page 225, 2016.

\bibitem{pointcheval2000security}
David Pointcheval and Jacques Stern.
\newblock Security arguments for digital signatures and blind signatures.
\newblock {\em Journal of cryptology}, 13(3):361--396, 2000.

\bibitem{ruohonen2018empirical}
Jukka Ruohonen.
\newblock An empirical analysis of vulnerabilities in python packages for web
  applications.
\newblock {\em arXiv preprint arXiv:1810.13310}, 2018.

\bibitem{santos2009towards}
Nuno Santos, Krishna~P Gummadi, and Rodrigo Rodrigues.
\newblock Towards trusted cloud computing.
\newblock {\em HotCloud}, 9(9):3, 2009.

\bibitem{schuh2015bitshares}
Fabian Schuh and Daniel Larimer.
\newblock Bitshares 2.0: General overview.
\newblock {\em accessed June-2017.[Online]. Available: http://docs. bitshares.
  org/\_downloads/bitshares-general. pdf}, 2015.

\bibitem{ProofOfAuthority}
Parity Tech.
\newblock Proof-of-authority chains - wiki.
\newblock \url{https://wiki.parity.io/Proof-of-Authority-Chains}, 2018.
\newblock Accessed: 2019-01-13.

\bibitem{archlinux}
Judd Vinet and Aaron Griffin.
\newblock Archlinux: A simple, lightweight distribution.
\newblock \url{https://www.archlinux.org/}, 2019.
\newblock Accessed: 2019-01-13.

\bibitem{archpackages}
Judd Vinet and Aaron Griffin.
\newblock Package search.
\newblock \url{https://www.archlinux.org/packages/}, 2019.
\newblock Accessed: 2019-01-13.

\bibitem{xu2016Blockchain}
Xiwei Xu, Cesare Pautasso, Liming Zhu, Vincent Gramoli, Alexander Ponomarev,
  An~Binh Tran, and Shiping Chen.
\newblock The blockchain as a software connector.
\newblock In {\em 2016 13th Working IEEE/IFIP Conference on Software
  Architecture (WICSA)}, pages 182--191. IEEE, 2016.

\bibitem{zheng2017overview}
Zibin Zheng, Shaoan Xie, Hongning Dai, Xiangping Chen, and Huaimin Wang.
\newblock An overview of blockchain technology: Architecture, consensus, and
  future trends.
\newblock In {\em Big Data (BigData Congress), 2017 IEEE International Congress
  on}, pages 557--564. IEEE, 2017.

\bibitem{githubit}
Felipe Zimmerle.
\newblock Blockchain repo.
\newblock \url{https://github.com/zimmerle}, 2019.
\newblock Accessed: 2019-01-13.

\end{thebibliography}

\end{document}